\documentclass[12pt]{article}

\usepackage{scicite}
\usepackage{amsmath}
\usepackage{graphicx}

\usepackage{times}

\newenvironment{sciabstract}{%
\begin{quote} \bf}
{\end{quote}}

\title{A spinwave Ising machine}

\author
{Artem Litvinenko$^{1\ast}$, Roman Khymyn$^{1}$, Victor H. González$^{1}$, Ahmad A. Awad$^{1}$, \\Vasyl Tyberkevych$^{2}$, Andrei Slavin$^{2}$ and Johan \AA kerman$^{1\ast}$\\
\\
\normalsize{$^{1}$Department of Physics, University of Gothenburg, Fysikgränd 3, 412 96 Gothenburg, Sweden}\\
\normalsize{$^{2}$Department of Physics, Oakland University, 48309, Rochester, Michigan, USA,}\\
\\
\normalsize{$^\ast$To whom correspondence should be addressed; E-mails:} \\
\normalsize{artem.litvinenko@physics.gu.se, johan.akerman@physics.gu.se}
}

\date{}

\begin{document} 


\baselineskip24pt


\maketitle 

\begin{sciabstract}
We demonstrate a spin-wave-based time-multiplexed Ising Machine (SWIM), implemented using a 5 $\mu$m thick Yttrium Iron Garnet (YIG) film and off-the-shelf microwave components. The artificial Ising spins consist of 34--68 ns long 3.125 GHz spinwave RF pulses with their phase binarized using a phase-sensitive microwave amplifier. Thanks to the very low spinwave group velocity, the 7 mm long YIG waveguide can host an 8-spin MAX-CUT problem and solve it in less than 4 $\mu$s while consuming only 7 $\mu$J. Using a real-time oscilloscope, we follow the temporal evolution of each spin as the SWIM minimizes its energy and find both uniform and domain-propagation-like switching of the spin state. 
The SWIM has the potential for substantial further miniaturization, scalability, speed, and reduced power consumption, and may become a versatile platform for commercially feasible optimization problem solvers with high performance.
\end{sciabstract}

\section*{Introduction}

As Moore’s law comes to an end due to physical limitations, while the amount processing data is constantly growing, novel analog and digital computing paradigm are being investigated. Large part of data processing tasks consist of combinatorial optimization problems and require special purpose accelerators for power efficient and fast computing. Combinatorial optimization is essential to select the optimal path from an enormous range of choices and appears in various social and industrial fields such as optimizing financial transactions\cite{Tatsumura2020}, logistics, including travel\cite{Zhang2021} and channel assignment for wireless communications\cite{wireless-allocation}, genetic engineering and molecular design for drug discovery\cite{combinatorialChemistry}. Conventional computers based on Von Neumann architectures are inefficient in solving hard combinatorial optimization problems due to the factorial growth of all possible combinations to be evaluated using brute force. Fortunately, Ising machines have emerged as a promising non-von-Neumann computing scheme that can accelerate computation of NP-hard optimization problems. 

An Ising machine maps an NP-problem onto the Ising Hamiltonian of an array of $N$ binarized and interacting physical entities, commonly referred to as spins $s_i=\pm 1$.
\begin{equation}
    H(s_1,...,s_N) =  -\sum_{i<j}J_{ij}s_i s_j - \sum_{i=1} h_i s_i
    \label{eq:ising-hamilt}
\end{equation}
The coding of a specific problem is performed by adjusting the pairwise coupling $J_{ij}$ terms such that the lowest energy spin glass configuration represents the solution to the NP-problem. This is possible because spin glass solutions belong to the NP-hard complexity class \cite{Barahona_1982}, and many NP-complete and NP-hard problems can be reformulated as Ising Hamiltonians \cite{lucas_ising_2014}. Then, the energy minimization of the entire array can be employed as a solution search algorithm in a process called annealing \cite{Mohseni2022}. 

Various types of Ising machines\cite{mohseni2022IMreview} have been proposed in the recent decades exploiting novel physical building blocks such as spintronic devices \cite{albertsson2021ultrafastSHNOIM, houshang2022SHNOIM, mcgoldrick2022SHNOIMs, Sutton2017}, memristor crossbars\cite{Cai2020}, quantum superconducting junctions\cite{johnson2011quantumIM,Bunyk2014}, metal-insulator relaxation oscillators \cite{dutta_experimental_2019}, and degenerate optical parametric oscillators \cite{Honjo2021SciAdv100kCIM,Inagaki2016Sci2000CIM,McMahon2016Sci100CIM} as well as employing conventional CMOS technology with analog electric oscillators \cite{Sutton2017, TWang2019}, FPGAs\cite{patel2020isingFPGA,tatsumura2021SBFPGA}. In particular, optical Coherent Ising Machines have attracted great attention due to their high computational speed, time-multiplexing method that provides all-to-all Ising spin connections and the largest amount of supported Ising spins amongst all other implementations. 

Optical CIMs use time-multiplexed degenerate optical parametric oscillators (DOPOs) that propagate in a ring cavity in the form of short light-wave pulses. All-to-all coupling between DOPOs can be implemented either physically, via external optical delay lines \cite{takata16bitCIMdelaylines2016,yamamoto2017CIMquantumfeedback,Haribara2016CIMPerformanceEvalDelayLines,marandi2014networkCIM4delaylines}, or electronically, using FPGA blocks \cite{McMahon2016Sci100CIM,Inagaki2016Sci2000CIM,yamamoto2017CIMquantumfeedback,Kako2020CIMwithErrFB}. In contrast, spatially-resolved IMs based on physical oscillators are inherently limited to low density connectivity as the number of all-to-all interconnections grows as $O(n^2)$ and hence represent an unsolvable problem for 2.5-D mass electronics production technology. DOPOs phase degeneracy allows one to exploit the bistable DOPOs phase as equivalent two-state Ising spins. CIM belongs to a annealing scheme class known as dynamical system solvers as their collective steady-state oscillation minimizes the individual phases described by the Kuramoto model.
\begin{equation}
    \dot{\Phi}_i=\omega_0 + K\sum_j J_{ij} \sin (\Phi_i-\Phi_j) +Kh_i \sin(\Phi_i-\omega_0 t)
     \label{eq:kuramoto}
\end{equation}
Where $K$ is a global coupling factor. See that eq.\ref{eq:kuramoto} reduces to eq.\ref{eq:ising-hamilt} for $\sin (\Phi_i-\Phi_j)=\pm 1$, promoting binarization which can be done by the parametric pumping at double frequency which will result in an additional term in eq.\ref{eq:kuramoto}. In the rotating frame, the steady-state solution ($\dot{\Phi}_i=0$) necessitates $\Phi_i-\Phi_j=0, \pi$ and $\Phi_i-\omega_0 t=0,\pi$ for $N$ mutually coupled oscillators. This state will have the properties of a spin glass solution and thus the architecture can be used as scalable NP-solvers, as shown in a progressive evolution from 100 to 100,000 optical spins in ring-based CIMs. 
Nevertheless, the commercial feasibility of optical CIMs remains elusive as the technology requires optical tables, kilowatts of power, and kilometers of optical fibers, which altogether blocks its further development from a proof-of-principle demonstration to a miniaturized commercially viable device.

In this work, we present a spinwave Ising machine (SWIM) with artificial spin states implemented via the phase of spinwave RF pulses propagating in an Yttrium iron garnet (YIG) thin film. Spinwaves are being actively used for implementation of signal processing \cite{papp2017SWspectrumAnalyzer,ustinov2010SWmultiSigProc,ustinov2008SWphaseShifter,ustinov2013NonlPhaseShift,sadovnikov2018SWdropFilter,grachev2022strainInterfer,ustinova2016SWinterfer,etko2011broadbandSWDL}, frequency synthesis \cite{litvinenko2018chaotic,litvinenko2021tunable,VitkoBistable2018}, logic \cite{ustinov2019SWlogic,nikitin2015SWlogic,Wang2020magnoniclogic,khitun2010magnoniclogic,khitun2011nonVolMagnLogic,khitun2008spinlogic,davies2015SWmultiplex} and computing \cite{Watt2021SWcomputer,watt2021SWReservoir,balinskiy2022primeComputing,khivintsev2022spinComputing,guo2021spintronicsComputation} applications due to their inherent non-linearities and exceptionally slow propagation velocities. In comparison to optical pulses, the use of spinwaves allows for 5--7 orders of magnitude reduction in delay line length. Another advantage comes from the simplicity and efficiency of microwave current-to-spinwave transducers and the off-the-shelf availability of power-efficient microwave electronic components and circuits for the signal processing required in a time-multiplexing architecture of Ising machines. In particular, it allows the use of electronic phase-sensitive amplification blocks consuming only milliwatts of power in contrast to optical kilowatt-power parametric phase-sensitive amplifiers. Altogether, the use of propagating spinwaves for the design of a time-multiplexed Ising machine brings the concept from a laboratory demonstrator to a commercially feasible technology with potential for CMOS integration.

\section*{Design and characteristics of the spinwave Ising Machine}

\begin{figure*}[hbt!]
\includegraphics[width=16cm]{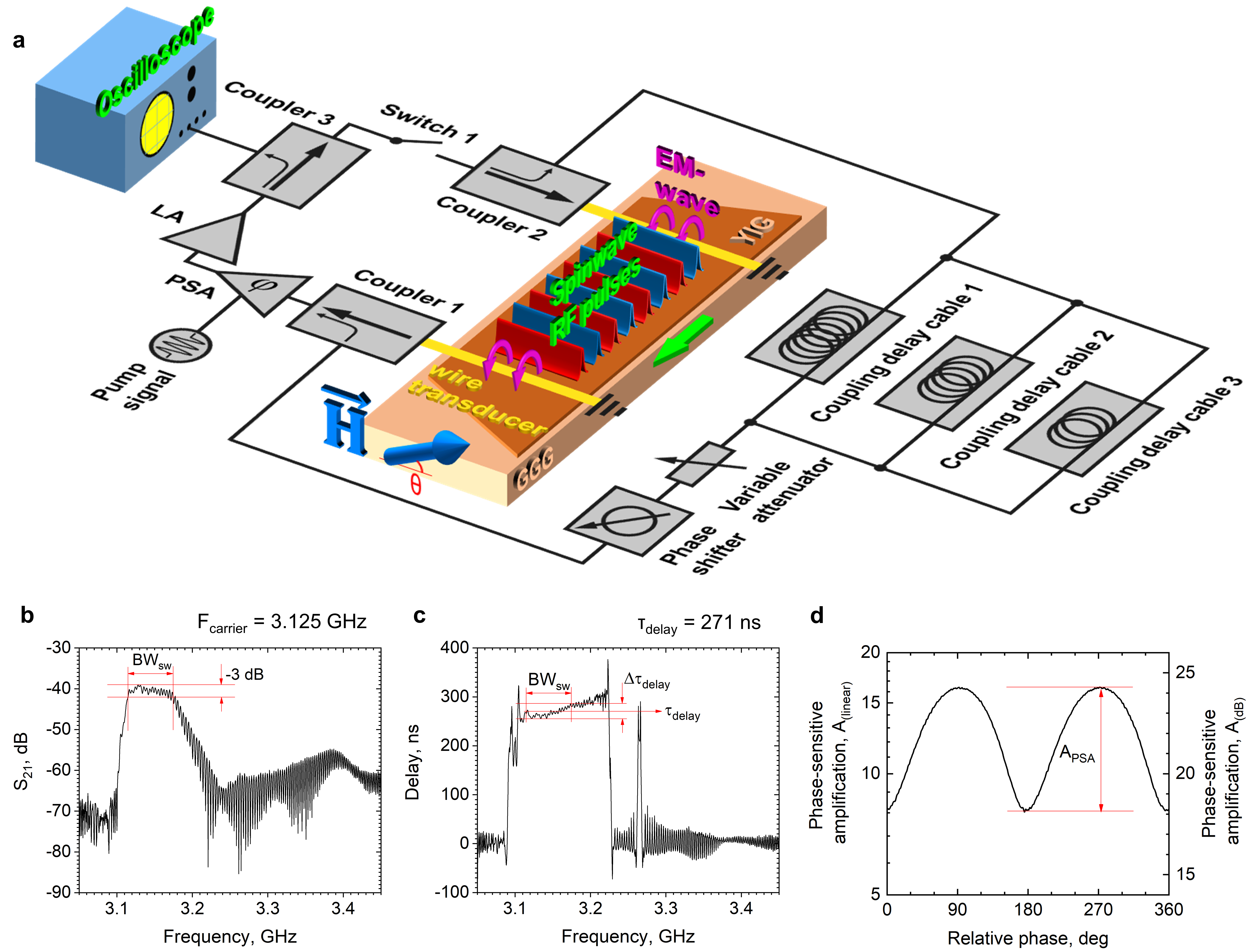}
\caption{\textbf{The spinwave Ising machine (SWIM).}
(\textbf{a}) Schematic of the SWIM with the in-plane magnetized YIG waveguide in the center together with peripheral microwave components, described in the text, and a real-time oscilloscope for direct studies of the time-evoluation of all spins; couplings between spins are implemented using three delay cables of different lengths. (\textbf{b}) Measurement of $S_{21}$ of the YIG delay line showing the spinwave spectrum. 
BW$_{sw}=$ 60 MHz is the spinwave spectral bandwidth measured at --3 dB.  
(\textbf{c}) The frequency-dependent delay time of the YIG delay line. $\tau_{delay}=$ 271 ns is the average delay time over 
BW$_{sw}$ with a total deviation of $\Delta\tau_{delay}=$ 30 ns.
(\textbf{d}) Amplification of the parametric phase-sensitive amplifier as a function of the relative phase between the propagating RF pulses and the reference signal $\omega_{ref}=$ 6.25 GHz, with a phase sensitivity of $\mathit{A_{PSA}}=$ 6.1 dB.
} 
\label{FigSWIMbasics} 
\end{figure*}

The SWIM employs a time-multiplexing approach similar to optical CIMs \cite{Honjo2021SciAdv100kCIM,Inagaki2016Sci2000CIM,McMahon2016Sci100CIM,yamamoto2017CIMquantumfeedback,Kako2020CIMwithErrFB,takata16bitCIMdelaylines2016,Haribara2016CIMPerformanceEvalDelayLines,marandi2014networkCIM4delaylines} but in the microwave frequency domain, which leads to certain signal processing modifications and substantial performance improvements. As shown in Fig.~\ref{FigSWIMbasics}, it consists of two blocks: a) an electronic part for linear and parametric amplification of the RF pulses, RF pulse interconnection, and RF pulse measurement, and b) a physical YIG spinwave waveguide where all the spinwave RF pulses are excited by a first transducer (antenna), allowed to propagate along the film length, and finally transformed back to a microwave RF pulses by a second transducer. In contrast to optical CIMs, where optical pulses propagate and are amplified in an optical ring system, never leaving the ring, the SWIM is a multi-physical system where amplification of the propagating RF pulses and their further signal processing is performed outside of the YIG waveguide in a power-efficient electronic system. This is possible because the carrier frequency of the propagating spinwaves is around 3 GHz and, therefore, can be easily handled by inexpensive and readily available commercial RF components.

Spin waves are a fundamental type of excitation in magnetic systems\cite{serga2010YIGMagnonics, cherepanov1993TheSagaOfYIG, Glass1988FerriteDevices,haidar2015thickness,navabi2019SWdamping}, representing collective precession of elementary magnetic moments coupled via exchange and dipole-dipole interactions. The spin-wave excitation frequency depends on the strength and orientation of the static magnetic field applied to the film and on the magnetic properties of its material, such as the saturation magnetization ($M_S$), the exchange stiffness ($A$), and gyromagnetic ratio ($\gamma$). In this work, the static magnetic field is applied at angle $\theta=$ 53$^\circ$ \emph{w.r.t.}~to the film plane, with its in-pane component parallel to the antennas, \emph{i.e.}~perpendicular to the spinwave propagation along the waveguide. In this configuration, the propagating waves are of a mixed type between magnetostatic surface spin waves (MSSW) and forward volume spin waves (FVMSW). The frequency in this configuration is described by the dispersion relation \cite{kalinikos1986theory, muralidhar2021femtosecond}: 
\begin{align}
    \omega_{0} &= \gamma \mu_0\sqrt{(H_{int}+M_s l_{ex}^2 k^2)(H_{int}+M_s l_{ex}^2 k^2+M_s F_0)},\\
    F_0 &= P_0+\cos^2\theta_{int}\left[1-P_0\right.+M_s\left.P_0(1-P_0)/\left(H_{int}+M_s l_{ex}^2k^2\right)\right],
    \label{eq:SWdispersionF}
\end{align}
where $k$ is the full wavevector of the spin wave, $P_0 = 1-(1-\exp(-kd))$ and $d=$ 5 $\mu$m being the thickness of the film, $l_{ex}= \sqrt{2 A/(\mu_0 M_s^2)}\simeq 1.7\cdot10^{-8}$ m for YIG, and $H_{int}$ and $\theta_{int}$ define the value and out of plane angle of the internal field; these can be found from the strength and angle of the applied field ($H$ and $\theta$) using the solution of the magnetostatic problem \cite{kalinikos1986theory, muralidhar2021femtosecond}. In our case $\mathit{H}=$ 0.04 T and the choice of $\theta = 53^\circ$ is motivated by an empirically found  high excitation efficiency and low total losses in the YIG delay line when using simple thin copper wire antennas. Moreover, spinwaves in such a configuration are strongly unidirectional and will be excited most efficiently in only one direction, which prohibits multi-transit spinwave echo signals and, therefore, improves the frequency stability of the SWIM.

Our time-multiplexed SWIM can be considered as a ring oscillator circuit. According to the circuit design theory, a circuit oscillates when the Barkhausen stability criteria are satisfied:
\begin{align}
        \beta \mathbf{A} &= 1,\\
        \angle \beta \mathbf{A} &= 2\pi n,\: n \in \{0,1,2...\} 
    \label{eq:BarkhausenStabCriteria}
\end{align}
where $\mathbf{A}$ is the total amplification in the loop, $\beta$ is the total loss, and $\angle \beta \mathbf{A}$ is the phase accumulated along the loop. The Barkhausen criteria are further narrowed by adding a parametric phase-sensitive amplifier (PSA), which limits stable oscillations only at either phase 0 or $\pi$ relative to a pumping reference signal at twice the oscillating frequency and consequently phase binarize the system. The PSA also induces second-harmonic frequency locking \cite{Lebrun2015PhaseNoiseSqueez,litvinenko2021analog} to the external reference signal,
\begin{align}
        \omega_{0}t-\omega_{ref}t/2 &= {\pi n},\: n \in \{0,1,2...\}\\
        \omega_{0} &= \omega_{ref}/2
    \label{eq:ExtBStabCriteria}
\end{align}
which further improves the frequency stability. 

These phase-binarized oscillation conditions are valid for continuous oscillations in the loop. In order to define separate time-multiplexed and phase-binarized aritificial Ising spins, we introduce an RF switch ($\bf{1}$) that is triggered by a square wave signal with variable frequency to control the spinwave RF pulse length, $\tau_p$. The frequency of the switching is determined by the total delay in the ring and the required number of supported artificial spins:
\begin{equation}
    f_{sw}=N/{\tau}_{delay}
     \label{eq:RFswitchFreq}
\end{equation}
The RF switch also prevents propagating spinwave RF pulses from spreading due to spinwave dispersion and non-constant delay time over the occupied frequency range.

One major advantage of using spinwaves is their 5--7 orders of magnitude slower propagation than the vacuum speed of light, which 
allows us to reduce the length of the YIG waveguide to millimeters compared to the kilometers-long optical fibers of CIM. 
As our artificial Ising spins consist of spinwave RF pulses of length $\tau_p$, the maximum number of supported spins is $N=\tau_{delay}/\tau_p$, where the waveguide delay time $\tau_{delay}=\mathit{l_{YIG}}/v_{g,sw}$ is the ratio of the waveguide length ($\mathit{l_{YIG}}$) and the spinwave group velocity ($v_{g,sw}$). An approximate theoretical value of the minimal pulse length 
is given by convolution of \emph{i}) the 3-dB bandwidth $BW_{sw}$ of the spinwave spectrum, which controls the slew rate of the spinwave RF pulses, and \emph{ii}) the delay time deviation $\Delta{\tau}_{delay}$ within $BW_{sw}$, which broadens the pulses as they propagate between the two transducers.

Fig.~\ref{FigSWIMbasics}(b) shows the spinwave spectrum of the YIG spinwave waveguide in the form of its $S_{21}$-parameter. The bandwidth of the spinwave generation spectrum is BW$_{sw}=$ 60 MHz measured at the --3 dB level. Fig.~\ref{FigSWIMbasics}(c) shows the corresponding frequency-dependent delay time, with an average value across BW$_{sw}$ of ${\tau}_{delay}=$ 271 ns and a deviation of 30.2 ns.The limit of minimal pulse width derived from bandwidth is 16.8 ns. However, the delay time deviation causes a significant broadening of propagating spinwave RF pulses limiting the total minimal pulse width by 34 ns. 

\section*{Defining and solving MAX-CUT problems}
In this section, we evaluate the computational performance of the SWIM on simple 4-spin and 8-spin MAX-CUT optimization problems. While we have successfully populated the waveguide with up to 10 spins, in agreement with the theoretical estimate above, we here limit the problem size to eight spins to avoid any undesired direct interaction between spinwave RF pulses as they begin to partially overlap when more closely packed. 

The first experiment demonstrates the principle of operation and routes for obtaining a solution to the simple 4-spin MAX-CUT problem with antiferromagnetic nearest-neighbor couplings described by the following Ising matrix:  

\begin{figure*}[hb!]
\includegraphics[width=16cm]{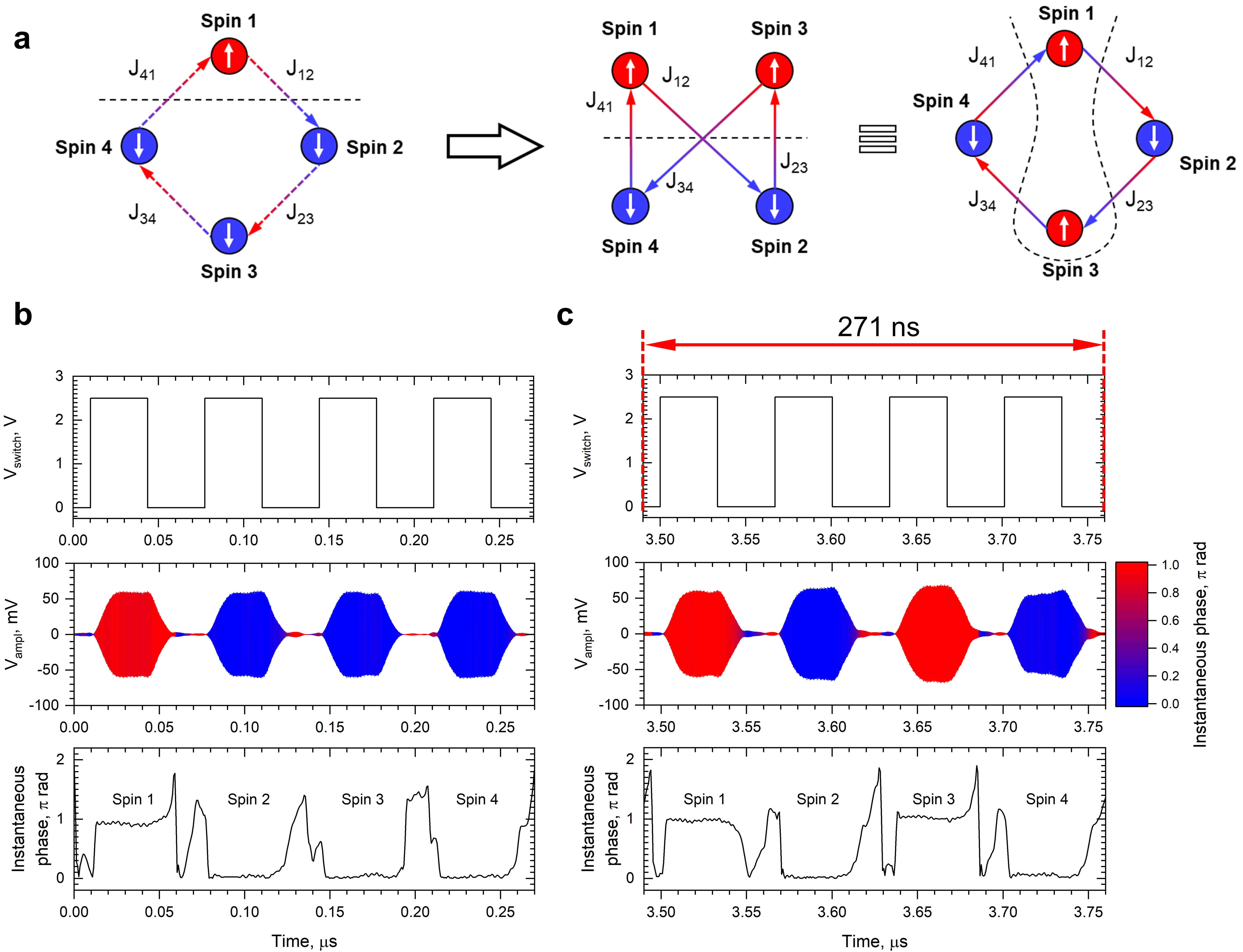}
\caption{\textbf{Experimental demonstration of 4-spin SWIM solving a MAX-CUT problem.}
(\textbf{a}) Transformation of the initial non-optimal state to an optimal solution for 4-spin MAX-CUT optimization problem. The number of cuts in each case is defined by the number of couplings which cross the line which separates spins if they were regrouped according to their spin values.
Time traces of SWIM control and computational signals in the initial non-optimal state (\textbf{b}) and the final solution (\textbf{c}). Top panel: a control signal of switch \textbf{1} with a repetition frequency of 14.8 MHz and duty cycle of 50$\%$ that forms oscillations in the SWIM ring oscillator into separate propagating RF pulses. Middle panel: an amplified signal of RF propagating pulses $V_{ampl}$. Bottom panel: a calculated instantaneous phase signal. The time traces in middle panels are colored according to a value of the calculated instantaneous signal to visually show the pulse instantaneous phase.}
\label{FigMAXCUT4} 
\end{figure*}
\begin{equation}
    J_{ij}=
    \left( \begin{array}{cccc}
    0 & -1 & 0 & 0 \\
    0 & 0 & -1 & 0 \\
    0 & 0 & 0 & -1 \\
    -1 & 0 & 0 & 0 
    \end{array} \right)
     \label{eq:MAXCUT4}
\end{equation}
This matrix can be realized using a single coupling delay cable ($\textbf{3}$) with a delay of 68 ns, \emph{i.e.}~of exactly one period of the pulse repetition time. Each microwave RF pulse that passes through Coupler $\textbf{2}$ generates both a spinwave RF pulse in the YIG delay line and a coupling microwave RF pulse propagating through the coupling delay cable $\textbf{1}$. After 68 ns, 
the coupling microwave RF pulse reaches Coupler $\textbf{3}$ and combines with a different RF microwave RF pulse converted from a propagating spinwave RF pulse, effectively implementing the coupling between these two spins. The negative sign of the coupling coefficients $J_{ij}$ is realized using an additional $180^\circ$ degree of phase shift implemented with a variable phase shifter. 

For a demonstration of the computational dynamics, the SWIM is first placed in a randomly chosen steady state $s_{j}^1= \{+1 -1 -1 -1\}$ with the coupling cable $\bf{3}$ disconnected. The value +1 corresponds to $180^\circ$ of phase difference between the RF pulse and the reference signal $\mathit{f_{ref}}$, while the value -1 corresponds to $0^\circ$ phase difference. The time traces of the signal of propagating RF pulses and their instantaneous phase are shown in Fig.~\ref{FigMAXCUT4}(b), where the color of the time trace corresponds to the instantaneous phase. As can be seen, the propagating spinwave RF pulses are well-defined and separated from each other; their individual phases are similarly well-defined and uniform in time. The initial $s_{j}^1$ represents a non-optimal MAX-CUT solution with the number of cuts equal to 2 (see Fig.~\ref{FigMAXCUT4}(a)\&(b)). At $t=0$ we connect the coupling cable $\bf{3}$ and let the SWIM evolve for 3.76 $\mu$s or 12 circulation periods, during which it reaches a stable state and finds the optimal number of cuts. Fig.~\ref{FigMAXCUT4}(c) shows the resulting spin state during the 12$^{th}$ circulation period, where the third spin has now switched its phase to $\pi$ under the influence of the coupling matrix. The SWIM has hence changed its state to $s_{j}^{12}= \{+1 -1 +1 -1\}$, which represents an optimal solution with the number of cuts equal to 4 (see Fig.~\ref{FigMAXCUT4}(a)). 

Similarly, we confirmed the SWIM's capability to solve 8-spin MAX-CUT problems. The frequency $f_{sw}$ of RF switch $\bf{1}$ was hence increased to 29.6 MHz according to eq.\ref{eq:RFswitchFreq} to increase the spin capacity to eight. 

As the corresponding repetition period decreased to 34 ns, a shorter coupling delay cable with 34 ns delay was used for this experiment. 
\begin{figure*}[hbt!]
\centering
\includegraphics[width=10cm]{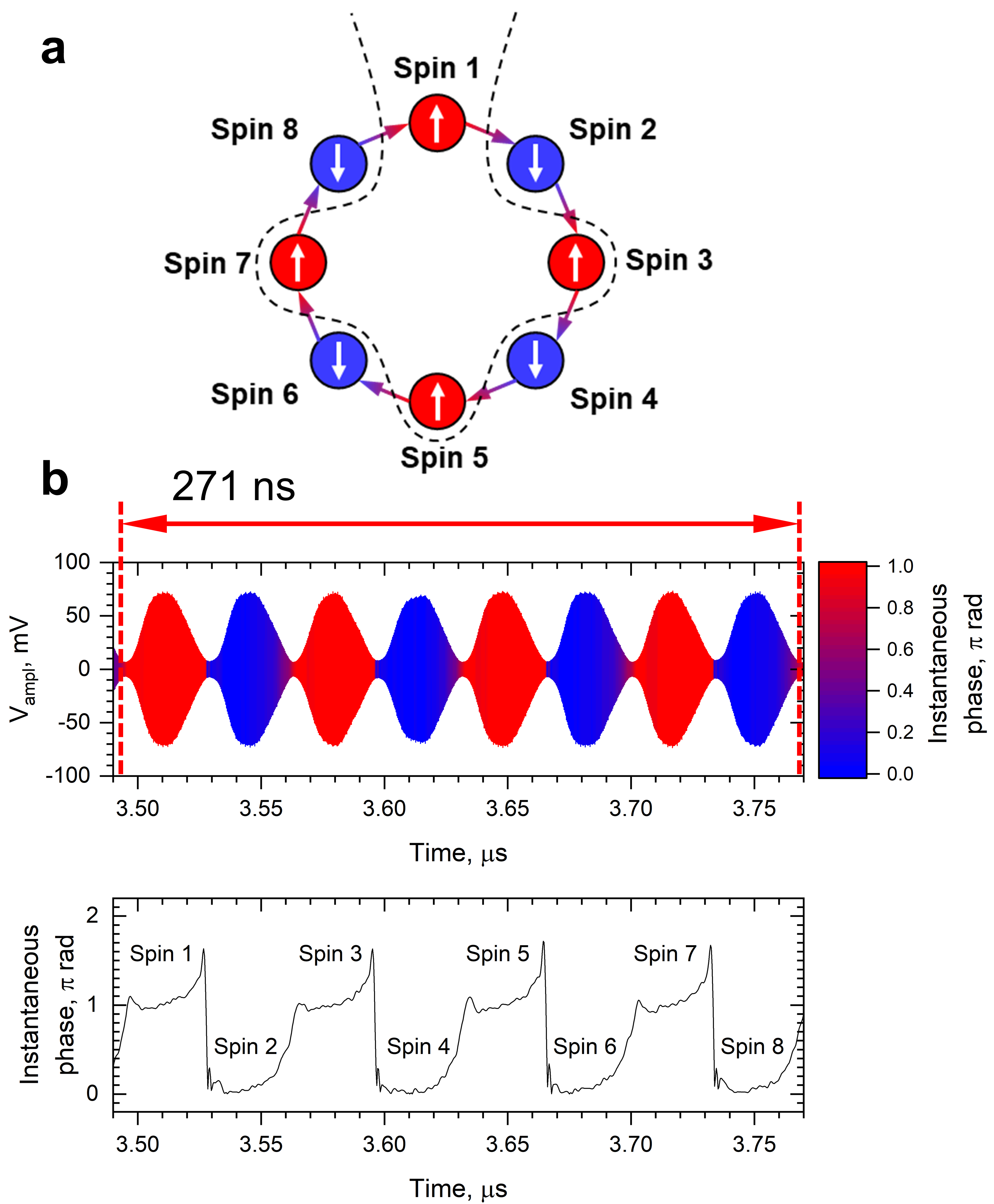}
\caption{\textbf{Experimental demonstration of 8-spin MAX-CUT problem computation.}
(\textbf{a}) An optimal solution for 8-spin MAX-CUT optimization problem with 8 cuts.
(\textbf{b}) Time traces of SWIM computational signals in the final optimal solution state. Top panel: an amplified signal of RF propagating pulses $V_{ampl}$. The total duration of the time trace signals corresponds to the delay time in YIG delay line. Bottom panel: a calculated instantaneous phase signal}
\label{FigMAXCUT8} 
\end{figure*}
As can be seen in Fig.~\ref{FigMAXCUT4}(b) the spinwave RF pulses start to show some minor overlap, which could nevertheless be neglected, as the SWIM had reached the correct solution when interrogated after 12 cycles.

\begin{figure*}[hb!]
\includegraphics[width=16cm]{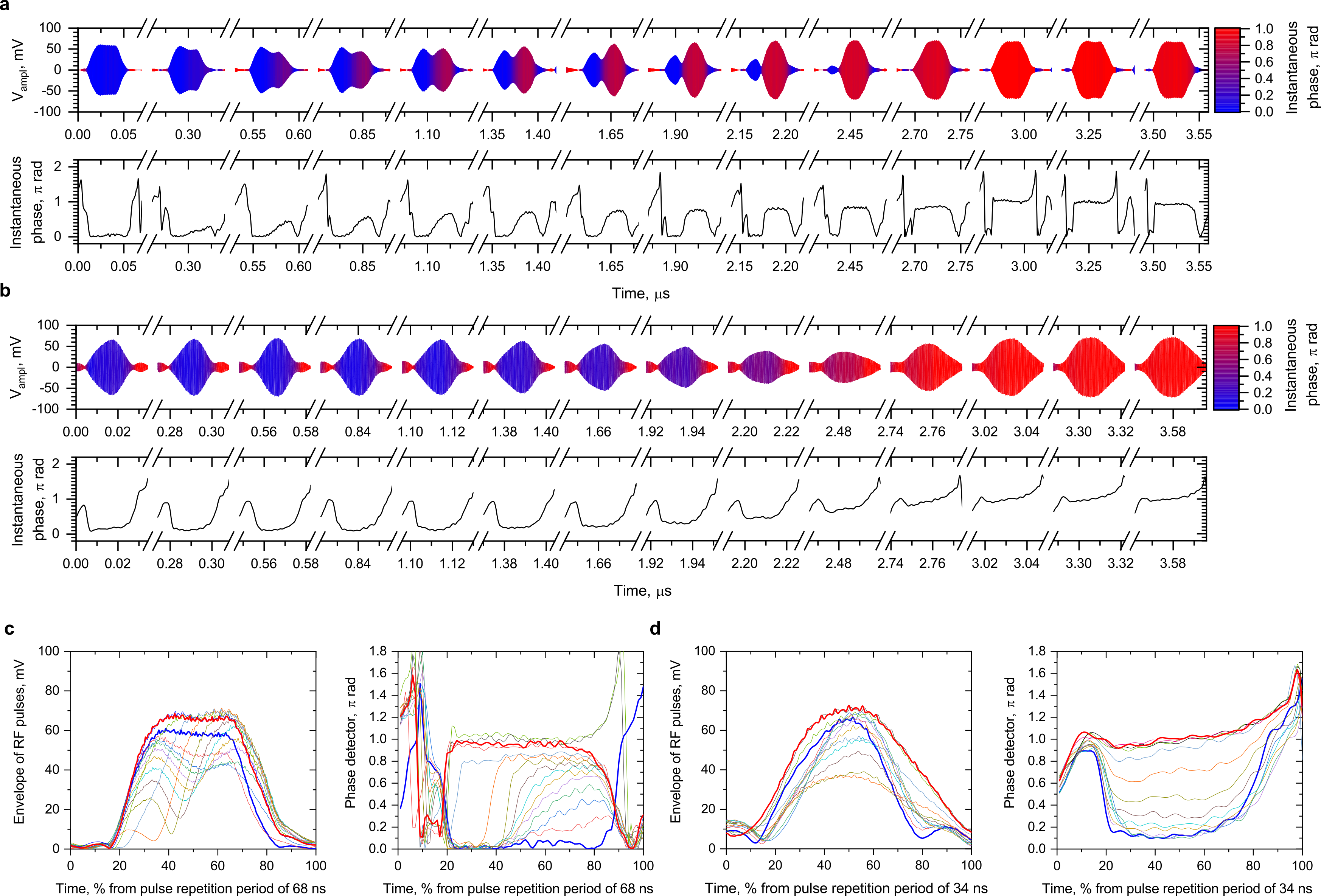}
\caption{\textbf{Spin switching scenarios for 4- and 8-spin MAX-CUT problem computations.}
(\textbf{a,b}) Evolution of the 3rd artificial spin from an initial non-optimal state to an optimal solution for 4-spin and 8-spin MAX-CUT optimization problem. Top panels: an amplified signal $V_{ampl}$ within time intervals which correspond to the 3rd propagating RF pulse. Bottom panels: a calculated instantaneous phase signal. 
(\textbf{c,d}) Envelopes (left panels) and instantaneous phase signals (right panels) of the 3rd propagating RF pulse signal within 12 circulation periods plotted in the form of overlapping traces in a relative scale. The color of the time traces corresponds to the circulation number.}
\label{figSpinEvolution} 
\end{figure*}

To further visualize the routes to a solution in greater detail, time traces and the instantaneous phases for the 3$^{rd}$ spin for both 4- and 8-spin MAX-CUT problems were captured at twelve consecutive circulation steps and compiled into Fig.~\ref{figSpinEvolution}(a,b) with scale brakes. The signal envelopes and instantaneous phases are also compiled in Fig.~\ref{figSpinEvolution}(c,d) in the form of overlapping traces with different colors corresponding to the number of circulations. Interestingly, the SWIM demonstrates two different scenarios of switching between $c_{3}^{1}=-1$ and $c_{3}^{12}=+1$. The first scenario is observed for the 4-spin MAX-CUT problem where the pulses are longer (40 ns). During the first 4 circulations, the spinwave RF pulse gradually separates into two distinct domains of approximately equal width, where only the trailing domain reverses its instantaneous phase. Withing the next 4-5 circulations the amplitude of the switching domain starts to decrease. The corresponding domain wall can also be clearly seen in the instantaneous phase. From this point on, the domain wall starts to propagate through the pulse until a fully switched single domain pulse is established at about the 10$^{th}$ cycle. The exact origin of this behavior is yet to be identified. 

A different switching mechanism can be observed for the 8-spin MAX-CUT  problem with the 34 ns long coupling delay cable $\bf{3}$. In this case, the evolution of the instantaneous phase occurs via a gradual and uniform shift from 0 to $\pi$. The difference between the two mechanisms is particularly clear when the evolutions of the envelope and instantaneous phase are plotted in Fig.4c\&d. It is possible that the shorter pulse length in the 8-spin case promotes a more uniform response to the coupling. 
Regardless of its origin, this gradual uniform switching mechanism might open up the possibility to use the SWIM concept as a potential platform to implement XY-Ising machines \cite{KalininXYIsing}.

A key parameter for benchmarking the performance of an Ising machine is its time-to-solution. In order to achieve optimal solutions in the simple cases of 4- and 8-spin MAX-CUT  problems, a single system run appears sufficient. However, for larger combinatorial problems with dense coupling graphs, there may exist a number of local minima in the Ising Hamiltonian. Therefore, a larger SWIM will require classical annealing schemes with multiple system runs and randomly chosen or idle initial conditions in order to perform statistical analysis on the probability of different stable solutions. The shorter the time-to-solution, the more such runs can be carried out in a given time. 

In our experiments, we periodically switch off the amplification in the loop for 2 $\mu$s to ensure that the SWIM returns to an idle state, after which we again switch on the loop amplification. Fig.~\ref{figSpinEvolution}(a) shows the temporal evolution of the SWIM prior to and after shutdown. It first takes 1.9 $\mu$s for the SWIM to develop circulating RF pulses to saturation level. It then takes an additional 1.5 $\mu$s to reach a stable state that represents the optimal solution. The rapid development of the amplitude of the circulating pulses is explained by a large small-signal over-amplification in the loop of the ring-based SWIM circuit ($\sim$6 dB) and leads to system evolution into a non-optimal solution before the signal reaches amplitude saturation. Reducing this over-amplification will lead to slower development of oscillation and it might improve the time-to-solution for large size optimization problems as at low amplitude of the RF pulses the SNR is lower and the system would more easily switch between its states. Another approach that could be attempted, similar to CIM  \cite{marandi2014networkCIM4delaylines}, is to gradually increase the amplification ratio until the system starts oscillating in the lowest energy state that corresponds to the minimum state of the Ising Hamiltonian.

\begin{figure*}[ht!]
\includegraphics[width=16cm]{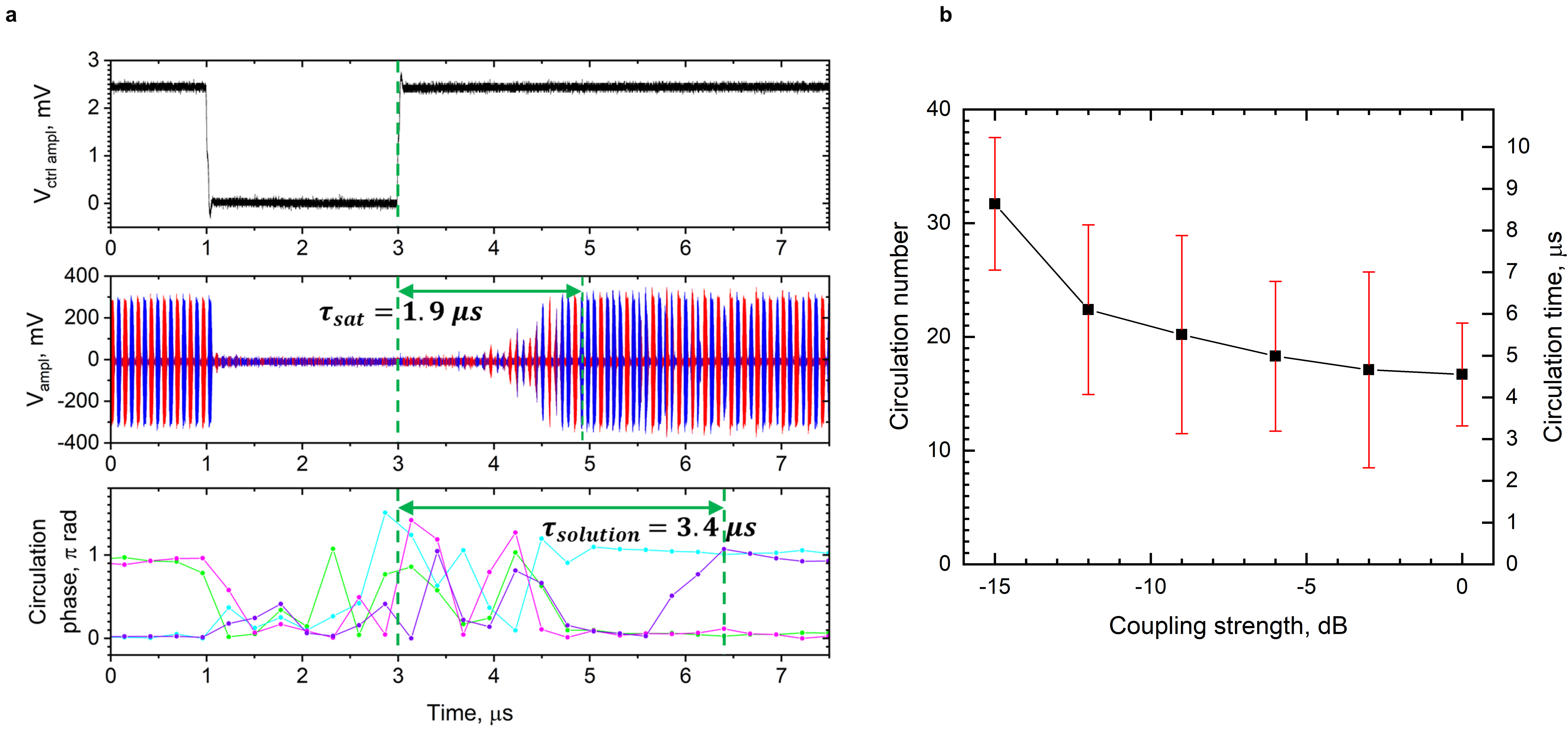}
\caption{\textbf{Measurement of the time-to-solution parameter for 4-spin MAX-CUT problem computation.}
(\textbf{a}) SWIM time traces for the measurement of time-to-saturation and time-to-solution parameters. Top panel: a control signal $V_{ctrl\:ampl}$ for the linear amplifier. Middle panel: an amplified signal of RF propagating pulses $V_{ampl}$. Bottom panel: phase values in the center of each propagating RF pulse sampled at each circulation period. The linear amplifier switches down at the moment $1\:\mu s$ with blackout period of $2\:\mu s$ to ensure signal suppression. 
(\textbf{b}) Circulation number and time-to-solution parameters for 4-spin MAX-CUT problem as a function of the coupling strength.}
\label{FigSpins4SwitchingStatistics} 
\end{figure*}

The SWIM design allows for the control of the overall coupling strength between spins by attenuating the signal coming from the coupling delay lines with a variable attenuator (see Fig.~\ref{FigSWIMbasics}(a)). To benchmark the SWIM dynamic range, a series of ten consecutive time-to-solution measurements was conducted at different coupling strengths for the 4-spin MAX-CUT problem. A statistical analysis is presented in Fig.~\ref{FigSpins4SwitchingStatistics}(b). For this very simple problem, the SWIM demonstrates almost constant time-to-solution values for coupling strengths from 0 to --12 dB with about 20 circulations or 5 $\mu$s. The very weak trend of a longer time-to-solution at weaker coupling strength is then more pronounced at --15 dB. The large SWIM dynamical range of 15 dB should allow the mapping of a wide range of optimization problems.

\section*{Discussion and outlook}
The SWIM concept has a high potential for further scaling in terms of spin capacity and physical size and the improvement of power consumption and speed performance parameters. Exploitation of non-linear spinwave solitons \cite{ustinov2008SWsolitons,ustinov2009SWsolitons,Kolodin1998SWsolitonAmpl,kostylev2000SWSolitonampl} and proper choice of magnetization angle could compensate spinwave dispersion and flatten delay time over a wide range of frequencies allowing to use shorter spinwave RF pulses effectively increasing the number of supported Ising spins with the same length of YIG waveguide. The use of exchange-dominated spinwaves with exceptionally slow propagation velocity \cite{serga2010YIGMagnonics,chumak2017magnonic,mahmoud2020SWcomputing,tikhonov2019spin} in nanoscale YIG films could help to further miniaturize the YIG waveguide.

The use of miniaturized microwave signal processing components opens a possibility of SWIM scaling by parallelizing the system using multiple ring circuits comprising identical phase sensitive amplifiers, YIG delay lines, etc. Parallelizing SWIM would improve the time-to-solution parameter by the number of the parallel rings as it decreases circulation time. Such a solution would require an FPGA\cite{McMahon2016Sci100CIM} as a highly parallel control system for the interconnection of Ising spins.

\section*{Conclusion}
We have demonstrated a spinwave time-multiplexed Ising machine (SWIM) and characterized its computational performance and functionality. We have shown how the SWIM can solve 4- and 8-spin NP-hard MAX-CUT problems in about 3.5 $\mu$s, which is comparable to D-wave's performance and faster than CIM. The multiphysics design allows the use of power-efficient and conventional low-power microwave components consuming 2 W of power, which amounts to an energy consumption of only 7 $\mu$J. This outperforms both D-wave's Advantage and CIM by 3--5 orders of magnitude. Thanks to the exceptionally slow spin wave propagation speed, the 7 mm long YIG waveguide can host up to 11 spins and further scaling is envisioned using both slower spinwave modes and patterned YIG waveguides. The vast variety of nonlinear spinwave modes and phenomena, and the direct availability of more optimized microwave signal processing components, bodes well for the SWIM scaling potential in terms of number of supported Ising spins, device size, and power consumption, making SWIM a commercially feasible platform for solving a wide range of optimization problems.
\bibliography{scibib}

\bibliographystyle{Science}

\section*{Author Contributions}
A.L. conceived the concept and designed the circuit; A.L., V.G., and A.A. performed the measurements and analyzed the data; A.L, R.K. performed theoretical calculations with help from A.S. and V.T.; J.Å. managed the project; all co-authors contributed to the manuscript, the discussion, and analysis of the results.

%\section*{Acknowledgments}
%Include acknowledgments of funding, any patents pending, where raw data for the paper are deposited, etc.

%\section*{Ethics declarations}A.L, R.K., and J.Å. are co-founders of a spin-off company that aims to develop spinwave-based Ising Machines. The rest of the authors declare no competing interests.

\end{document}